\documentclass[preprint,showpacs,amsfonts]{revtex4}
\usepackage{epsfig}

\def\ra{\rangle}
\def\la{\langle}

\begin{document}

\title{Monogamous property of generalized W states in three-qubit systems in terms of
relative entropy of entanglement}
\author{Zhen Wang}
\affiliation {Department of Mathematics, Jining University,
Qufu 273155, China\\
wangzhen061213@sina.com}
\author{Xiu-Hong Gao}
\affiliation {School of Mathematical Sciences, Capital Normal
University, Beijing 100048, China\\
gaoxiuh@cnu.edu.cn}
\author{Zhi-Xi Wang}
\affiliation {School of Mathematical Sciences, Capital Normal
University, Beijing 100048, China\\
wangzhx@cnu.edu.cn}

\begin{abstract}
Because of the difficulty in getting the analytic formula
of relative entropy of entanglement, it becomes troublesome to study the monogamy relations
of relative entropy of entanglement for three-qubit pure states. However, we find
that all generalized W states have the monogamous property for relative entropy of entanglement by
calculating the relative entropy of entanglement for the reduced states of the generalized W states in three-qubit systems.

\noindent{\it Keywords}: relative entropy of entanglement, generalized W states, monogamous property
\end{abstract}

\pacs{03.67.Mn, 03.65.Ud}

\maketitle

Entanglement has played a significant role in the field of
quantum information and quantum computation \cite{horodeckii}. This attracts an increasing interest
in the study of quantification of entanglement for any quantum state. Lots of entanglement
measures have been proposed for last two decades, e.g. entanglement of formation (EoF),
concurrence, distillable entanglement and relative entropy of entanglement (REE).
Among them, one of the important entanglement measures is the distillable entanglement
as it is the optimal rate at which one can extract maximally entangled states out of the given state.
However, it is not a trivial task to give the analytical expression of distillable entanglement
for general quantum states. Fortunately, the REE has been shown to provide the tight upper bound on distillable entanglement \cite{vedral1,vedral2}.

The REE can be used as a
distance function on the set of density operators. It is defined as
\begin{eqnarray}\label{def}
E_r(\rho)=\min_{\sigma\in{\mathcal D}}S(\rho\|\sigma)
=\min_{\sigma\in{\mathcal D}}{\rm tr}(\rho\ln\rho-\rho\ln\sigma),
\end{eqnarray}
where $\mathcal D$ is the set of separable states.
Although the analytical solutions have been given for the states of some special sets
with highly symmetry \cite{vedral1,vedral2,vollbrecht,rains,chen,wei,miranowicz,wang,park},
it is still an open fundamental problem \cite{krueger} to obtain a closed formula for a two-qubit state
due to including the convex optimization problem in terms of REE. Recently, for a given closest separable
state (CSS) $\sigma$ in two-qubit system,   Miranowicz and Ishizaka have given the REE of all entangled
states $\rho$ which have $\sigma$ as their CSS and obtained the analytical solution of REE in some special cases \cite{miranowicz}.
In fact, Friedland and Gour have found that there exists the closed formula of all entangled states for a given
CSS on the boundary of separable states in \cite{gour}.  In \cite{miranowicz,miranowicz2},
Kim {\it et. al.} studied the inverse process and found that
it is still an unsolved problem.

On the other hand, multipartite entanglement plays an important role in condensed matter physics.
The seminal work of Coffman, Kundu
and Wootters (CKW)\cite{ckw} provided a way to quantify the three-party entanglement by introducing
the three-tangle. Since then, many researchers began to address the monogamy relations of
entanglement \cite{ou,yu,kim,zhao,adesso,corenelio}. The monogamy inequality
in terms of concurrence in three-qubit
systems introduced
by CKW:
\begin{equation}\label{ckw}
C^2_{AB}+C^2_{AC}\leq C^2_{A:BC},
\end{equation}
where $C^2_{AB}$ ( $C^2_{AC}$) is the concurrence between $A$ and $B$ ($C$), while $C^2_{A:BC}$ ---between system $A$
and $BC$.  Recently, Osborne and
Verstraete \cite{osborne} have generalized the CKW inequality to
$n$-qubit systems.
In addition, the monogamy inequality holds in terms of distillable
entanglement \cite{bennett}, negativity \cite{vidal} and squashed
entanglement \cite{chri}, but fails for other definitions such as
 the concurrence itself and the EoF
\cite{wootters,bennett} for a three-qubit state. This indicates that
it is of crucial important for the
proper choice of the entanglement measure to capture
the monogamous nature of
quantum entanglement. Now one may ask
whether the monogamy inequality holds in terms of REE for a
three-qubit pure state. As it is unsolved to find the analytical
expression for a two-qubit state in terms of REE \cite{miranowicz,miranowicz2}, the monogamy relations
of REE for three-qubit states become very troublesome problem.

It is known that there are two inequivalent classes of genuine tripartite entangled states
characterized by means of local operations and
classical communication (LOCC) for three-qubit
states \cite{dur}. One is the Greenberger-Horne-Zeilinger (GHZ) class
\cite{green89}, and the other is the W class \cite{dur}. The feature of GHZ-class states is that when any
one of the three qubits is traced out, the remaining two are separable. Hence, all GHZ-class states are
monogamous for any entanglement measure. The W-class states have the property that their
entanglement has the highest degree of endurance against loss of one of the three qubits. Thus
it is worth investigating the monogamous property of W-class states. In this paper, we only investigate
the monogamy relations of generalized W states
\begin{equation}\label{w}
|\psi_W\ra=\alpha|001\rangle+\beta|010\rangle+\gamma|100\rangle,
\end{equation}
where $\alpha, \beta, \gamma \in {\bf R}$ and $\alpha^2+\beta^2+\gamma^2=1$.

It is found that the reduced states of the states (\ref{w})
have the following form:
\begin{equation}\label{partial}
\rho=\left(
\begin{array}{cccc}
a&0&0&0\\
0&b&\sqrt{bc}&0\\
0&\sqrt{bc}&c&0\\
0&0&0&0\\
\end{array} \right),
\end{equation}
where $a+b+c=1$.
Moreover, the states with the form (\ref{partial}) are
$U_{\bf x}\otimes U_{\bf x}$-invariant states \cite{chru}.
Recall that the most general two-qubit state which is
$U_{\bf x}\otimes U_{\bf x}$-invariant has the following form \cite{chru}:
\begin{equation}\label{ux2}
\rho=\left(
\begin{array}{cccc}
a&0&0&0\\
0&b&f&0\\
0&f&c&0\\
0&0&0&d\\
\end{array} \right),
\end{equation}
where $a+b+c+d=1$ and $bc\geq f^2$.
Accordingly, the most general state which is $U_{\bf x}\otimes
U_{\bf x}^*$-invariant state has the form:
\begin{equation}\label{uxy2}
\rho^\prime=\left(
\begin{array}{cccc}
b&0&0&f\\
0&a&0&0\\
0&0&d&0\\
f&0&0&c\\
\end{array} \right),
\end{equation}
where $a+b+c+d=1$ and $bc\geq f^2$.

At first, we derive the REE of reduced states of generalized W states in three-qubit systems
by use of their symmetric property.

{\bf Theorem}\quad  For $\rho$ with the form (\ref{partial}), we have
\begin{eqnarray}\label{e1}
E_r(\rho)&=& a\log  a+2(1-a)\log (1-a) \nonumber\\[2mm]
&&+\log  [(1+M)(1+N)]-(b+c)\log (b M+2\sqrt{b c M N}+c N),
\end{eqnarray}
where
\begin{eqnarray}\label{ab}
M&=\displaystyle\frac{\sqrt{\Delta}+b-c-2a^2b}{2ab(1+a)},\nonumber\\
N&=\displaystyle\frac{\sqrt{\Delta}-b+c-2a^2c}{2ac(1+a)},
\end{eqnarray}
with $\Delta=(b-c)^2+4a^2bc.$

{\bf Proof.}\quad In order to prove the theorem we utilize the fact that
$$E_r(\rho)=\inf\{S(\rho\|\sigma)\big|\sigma\in\mathbf P\mathcal
D\}$$
when $\rho=\mathbf P\rho$, where $\mathbf P$ denotes a projector
projecting an arbitrary state onto the class (\ref{partial})
\cite{vollbrecht}. Moreover, a state from (\ref{ux2}) is separable iff it has the form:
\begin{equation}\label{separable}
\sigma^*=\left(
\begin{array}{cccc}
x&0&0&0\\
0&u&re^{i\theta}&0\\
0&re^{-i\theta}&v&0\\
0&0&0&y\\
\end{array} \right)
\end{equation}
with $u\geq0,\ v\geq0,\ x\geq0,\ y\geq0,\ uv-r^2\geq0,\  x+u+v+y=1$ and $xy-r^2\geq0$ by use of
PPT criterion \cite{peres}. Because the entangled state $\rho$ is not full rank, $\sigma^*$ must be
on the boundary of $\mathcal D$ \cite{gour}. Therefore, either $\sigma^*$ or its partial transposition $(\sigma^*)^{T_A}$
has at least one zero eigenvalue. Aussume $(\sigma^*)^{T_A}$ is singular.
The eigenvalues
$u,\ v,\ \frac{x+y+\sqrt{(x-y)^2+4r^2}}{2},\ \frac{x+y-\sqrt{(x-y)^2+4r^2}}{2}$
of $(\sigma^*)^{T_A}$
show that $(\sigma^*)^{T_A}$ is rank deficient when $r^2=xy$. Thus the CSS for
$\rho$ is the state (\ref{separable}) with $r^2=xy$. Suppose $uv-r^2=\epsilon$, where $\epsilon\geq0$, we
have $xy=uv-\epsilon$.
Thus
\begin{eqnarray}
E_r(\rho)&=&a\log  a+(b+c)\log (b+c)+\min\{-{\rm tr (\rho\log \sigma^*)}\} \nonumber\\[2mm]
&\geq&a\log  a+(1-a)\log (1-a)+\min\{-a\log \la00|\sigma^*|00\ra-(b+c)\log \la\psi|\sigma^*|\psi\ra
\nonumber\\[2mm]
&=&a\log  a+(1-a)\log (1-a)\nonumber\\[2mm]
&&\quad +\min\{-a\log  x-(b+c)
\log [\frac{b}{b+c}u+2\frac{\sqrt{b c x y}}{b+c}\cos\theta+\frac{c}{b+c}v]\}\nonumber\\[2mm]
&=&a\log  a+2(1-a)\log (1-a)\nonumber\\[2mm]
&&\quad +\min\{-a\log  x-(1-a)
\log [b u+2\sqrt{bc}\sqrt{uv-\epsilon}\cos\theta+c v]\},
\end{eqnarray}
where $|\psi\ra=(0,\ \sqrt{\frac{b}{b+c}},\ \sqrt{\frac{c}{b+c}},\ 0)^T$
and we have used the fact that $f(x)=-\log  x$ is convex. Now we have to minimize the function
$$f(\theta,\ \epsilon)=-a\log  x-(1-a)\log [bu+2\sqrt{bc}\sqrt{uv-\epsilon}\cos\theta+cv].$$
It is easy to
show that $f(\theta,\ \epsilon)$ is an increasing function on $\epsilon$ and have the minimization when
$\cos\theta=1$. Thus it is found that
$$f(\theta,\ \epsilon)_{\min}=-a\log  x-(1-a)\log [bu+2\sqrt{bcuv}+cv]$$
and $xy=uv$. Let
$$
x=\cos^2\theta_1\cos^2\theta_2,\ y=\sin^2\theta_1\sin^2\theta_2, \
u=\sin^2\theta_1\cos^2\theta_2,\ v=\cos^2\theta_1\sin^2\theta_2.
$$
The minimization problem of $E_r(\rho)$ has been translated into solving the minimum value of
\begin{eqnarray*}
g(\theta_1,\ \theta_2)&=&a\log  a+2(1-a)\log (1-a)-a\log \cos^2\theta_1\cos^2\theta_2\\
&&\hspace{5cm} -(1-a)\log [\sqrt{b}\sin\theta_1\cos\theta_2
+\sqrt{c}\cos\theta_1\sin\theta_2]^2.
\end{eqnarray*}
The explicit calculation of the minimization of
$E_r(\rho)$ is now a tedious
but straightforward exercise, whose result is quoted in the right
side of the Eq. (\ref{e1}). Therefore the lower limit of $E_r(\rho)$
can be reached. In addition, if $\sigma^*$ is singular, one can obtain the same
result by use of a similar analysis, which proves the Eq. (\ref{e1}). $\hfill\Box$

{\bf Example}\quad Set $a=1-\lambda,\ b=c=\frac{\lambda}{2}$ when $b=c$.
The state (\ref{partial}) becomes the most general example of  Vedral-Plenio states
$$\rho_{vp}=(1-\lambda)|00\ra\la00|+\lambda|\Psi^+\ra\la\Psi^+|,$$
where $|\Psi^+\ra=\frac{1}{\sqrt{2}}(|01\ra+|10\ra).$ By a calculation, it is found that
$M=N=\frac{\lambda}{2-\lambda}$ and the known result
$$E_r(\rho_{vp})=
(1-\lambda)\log (1-\lambda)+(\lambda-2)\log (1-\frac{\lambda}{2})$$
derived in \cite{vedral1,vedral2}.

{\bf Remark}\quad It is easy to show that the REE of the states (\ref{ux2}) and (\ref{uxy2})
are the same because they are local unitary equivalent. Thus one can obtain the corresponding
REE of $U_{\bf x}\otimes U^*_{\bf x}$-invariant states
\begin{equation}
\rho^\prime=\left(
\begin{array}{cccc}
b&0&0&f\\
0&a&0&0\\
0&0&0&0\\
f&0&0&c\\
\end{array} \right).
\end{equation}

\begin{figure}[tbp]
\begin{center}
\resizebox{8cm}{!}{\includegraphics{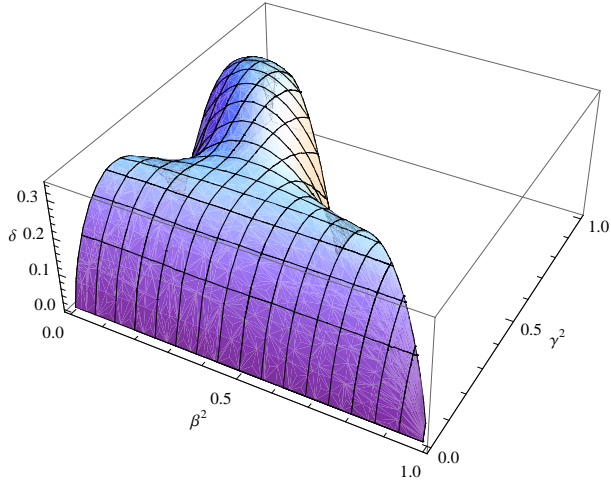}}
\end{center}
\caption{the image of $\delta$ \label{figure}}
\end{figure}

In the following we discuss the monogamy relations of the REE for the generalized W
states (\ref{w}). Clearly, the reduced density matrices of a
generalized W state (\ref{w}) can be written as:
\begin{equation}\label{ab}
\rho_{AB}={\rm tr}_C(|\psi_W\ra\la\psi_W|)=\left(
\begin{array}{cccc}
\alpha^2&0&0&0\\
0&\beta^2&\beta\gamma&0\\
0&\beta\gamma&\gamma^2&0\\
0&0&0&0\\
\end{array} \right),
\end{equation}
and
\begin{equation}\label{ac}
\rho_{AC}={\rm tr}_B(|\psi_W\ra\la\psi_W|)=\left(
\begin{array}{cccc}
\beta^2&0&0&0\\
0&\alpha^2&\alpha\gamma&0\\
0&\alpha\gamma&\gamma^2&0\\
0&0&0&0\\
\end{array} \right).
\end{equation}
For convenience, let $E^r_{AB}$, $E^r_{AC}$ and $E^r_{A:BC}(|\psi_W\ra)$
represent the REE of $\rho_{AB}$, $\rho_{AC}$ and a bipartite state of
(\ref{w}) on subsystems $A$ and $BC$, respectively. Moreover, denote
\begin{equation}\label{delta}
\delta=E^r_{A:BC}(|\psi_W\ra)-E^r_{AB}-E^r_{AC},
\end{equation}
where $$E^r_{A:BC}(|\psi_W\ra)=-\gamma^2\log \gamma^2-(1-\gamma^2)\log (1-\gamma^2).$$
In order to study the monogamy relations of generalized W states by REE, we consider the image of $\delta$
presented in FIG. \ref{figure}.

The FIG.1 shows that all generalized W states in three-qubit systems satisfy the monogamy relations of
REE. Thus the entanglement $A$-$BC$, measured by REE, is not completely determined by its partial entanglements,
$A$-$B$ and $A$-$C$. It shows the fact that the entanglement of these states
is very robust against particle loss \cite{higuchi}. Recently, the monogamy property of the quantum
correlations in three qubit pure states is considered.
Prabhu {\it et.al.} considered the monogamy property of quantum correlations by use of the quantum discord in \cite{Prabhu}.
It is found that generalized W states do not follow monogamy relation in terms of quantum discord.
In addition, monogamy of quantum correlations for generalized W states
using Rajagopal-Rendell quantum deficit is also addressed in \cite{sudha}. It is showed that
generalized W states exhibit both mono and polygamous nature unlike that of the W state. It is different from
the monogamy property for generalized W states in terms of REE.

In conclusion, it is found that all generalized W states have the monogamous property in terms of REE.
The result could shed new lights on the study of genuine multipartite entanglement.
In addition, we have obtained the analytic formula for the reduced states of generalized W states by their symmetric property.
In fact, these states are rank-2 two-qubit $U_{\bf x}\otimes U_{\bf x}$-invariant states. The investigation of
monogamous property of generalized W states is an application of expressions of their REE.
In \cite{dur}, it is shown that there are six different classes for three-qubit pure states
under LOCC with nonzero probability. Although it is
still not known if the monogamy relations in terms of REE hold for all three-qubit states, they
are true for all generalized W states and other five classes except W-class. After all, as it is shown that
a closed formula of REE for two qubits can be given only in some special cases in \cite{miranowicz,miranowicz2},
it is difficult to examine the monogamous property of all three-qubit states.

\bigskip
\noindent{\bf Acknowledgments}\, The project is supported by the Shandong Province Doctoral Fund (BS2010DX004) and
Jining University Youth Research Fund (2010QNKY08).

\end{document}